# Clinical Concept Extraction: a Methodology Review

Sunyang Fu[ac], David Chen[a], Huan He[a], Sijia Liu[a], Sungrim Moon[a], Kevin J Peterson[bc], Feichen Shen[a], Liwei Wang[a], Yanshan Wang[a], Andrew Wen[a], Yiqing Zhao[a], Sunghwan Sohn[a], Hongfang Liu[ac]

[a]Department of Health Sciences Research, Mayo Clinic, 200 First Street SW, Rochester, MN 55905

[b]Department of Information Technology, Mayo Clinic, 200 First Street SW, Rochester, MN 55905

[c]University of Minnesota – Twin Cities, Minneapolis, MN 55455

**Email Addresses:**
Sunyang Fu: Fu.Sunyang@mayo.edu
David Chen: Chen.David@mayo.edu
Huan He: He.Huan@mayo.edu
Sijia Liu: Liu.Sijia@mayo.edu
Sungrim Moon: Moon.Sungrim@mayo.edu
Kevin J Peterson: Peterson.Kevin@mayo.edu
Feichen Shen: Shen.Feichen@mayo.edu
Liwei Wang: Wang.Liwei@mayo.edu
Yanshan Wang: Wang.Yanshan@mayo.edu
Andrew Wen: Wen.Andrew@mayo.edu
Yiqing Zhao: Zhao.Yiqing@mayo.edu
Sunghwan Sohn: Sohn.Sunghwan@mayo.edu
Hongfang Liu: Liu.Hongfang@mayo.edu

**Corresponding Author:** Dr. Hongfang Liu, Division of Digital Health Sciences, Mayo Clinic, 200 First St SW, Rochester, MN 55905 (email: liu.hongfang@mayo.edu; phone: 507-293-0057)

**ABSTRACT**

**Background**

Concept extraction, a subdomain of natural language processing (NLP) with a focus on extracting concepts of interest, has been adopted to computationally extract clinical information from text for a wide range of applications ranging from clinical decision support to care quality improvement.

**Objectives**

In this literature review, we provide a methodology review of clinical concept extraction, aiming to catalog development processes, available methods and tools, and specific considerations when developing clinical concept extraction applications.

**Methods**

Based on the Preferred Reporting Items for Systematic Reviews and Meta-Analyses (PRISMA) guidelines, a literature search was conducted for retrieving EHR-based information extraction articles written in English and published from January 2009 through June 2019 from Ovid MEDLINE In-Process & Other Non-Indexed Citations, Ovid MEDLINE, Ovid EMBASE, Scopus, Web of Science, and the ACM Digital Library.

**Results**

A total of 6,686 publications were retrieved. After title and abstract screening, 228 publications were selected. The methods used for developing clinical concept extraction applications were discussed in this review.

# Graphical abstract

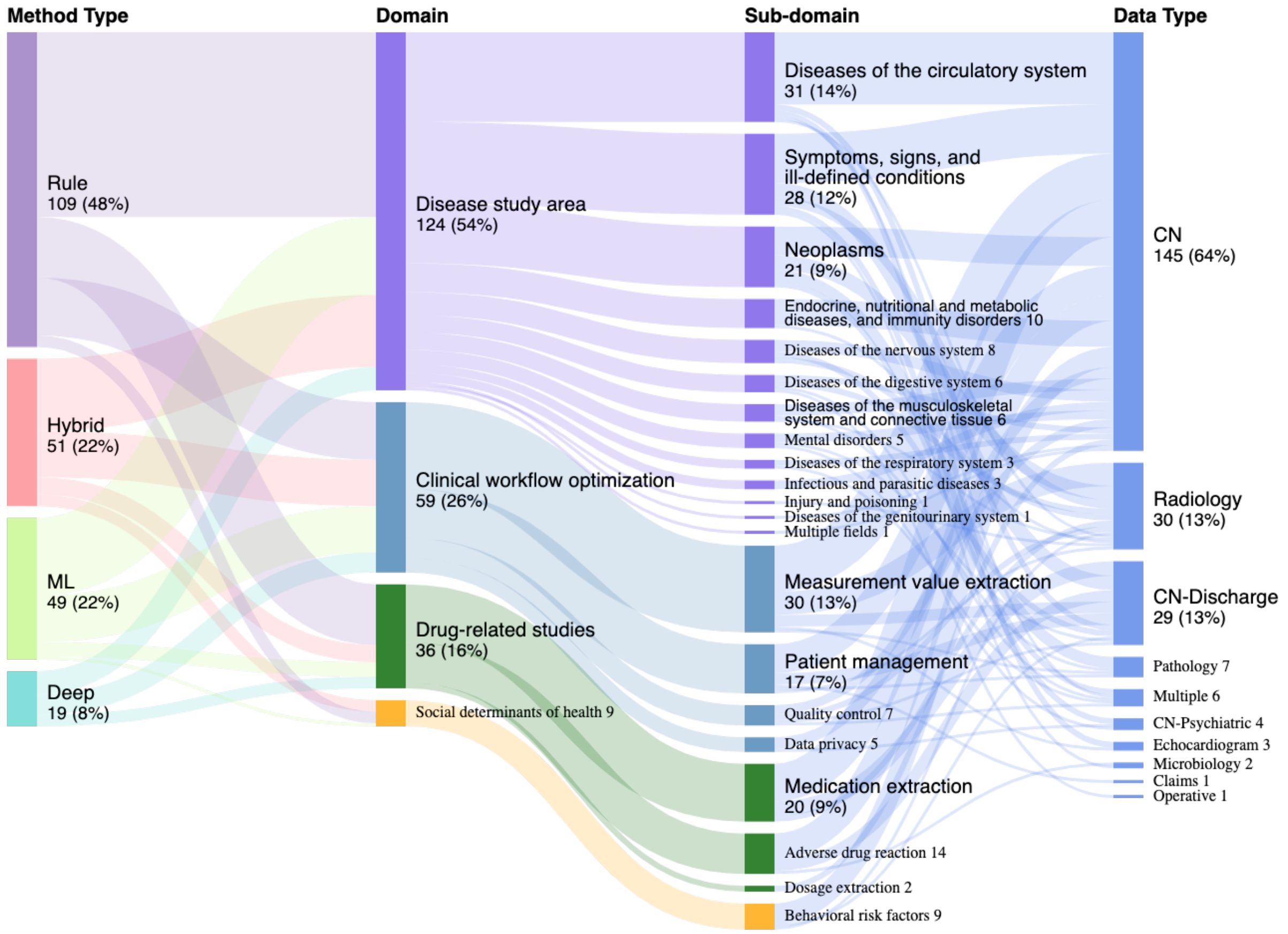

# 1. Introduction

Electronic health records (EHRs) have been widely viewed to have great potential to advance clinical research and healthcare delivery (1). To achieve the goal of "meaningful use", transforming routinely generated EHR data into actionable knowledge requires systematic approaches (2). However, a significant portion of clinical information remains locked away in free text (3). The solution to this is to replace these manual methods with autonomous computational extraction. The field of research related to this topic, commonly referred to as information extraction (IE), a subdomain of natural language processing (NLP), seeks to address the challenges of computationally extracting information from free-text narratives (4, 5). Specifically, we define the process of automatically extracting pre-defined clinical concepts from unstructured text as clinical concept extraction, which includes concept mention detection and concept encoding. Concept mention detection generally adopts the named entity recognition (NER) (6, 7) technology in the general domain, which focuses on detecting concept mentions in the text. Concept encoding aims to map the mentions to concepts in standard terminologies or those defined by downstreaming applications (5, 8, 9). Concept extraction has been adopted to extract clinical information from text for a wide range of applications ranging from supporting clinical decision making (3) to improving the quality of care (10). A review done by Meystre et al. in 2008 observed an increasing utilization of NLP in the clinical domain and a major challenge in advancing clinical NLP due to the unavailability of a large amount of clinical text (11). A summary of EHR-based clinical concept extraction applications can be found in Wang et al (5).

For illustrative purposes, an example of a typical clinical concept extraction task is presented in Figure 1, where the goal is to identify patients with silent brain infarcts (SBIs) from neuroimaging reports for stroke research. Silent brain infarcts, defined as old infarcts found through neuroimaging exams (CT or MRI) without a history of stroke, have been considered a major priority for new studies on stroke prevention by the American Heart Association and American Stroke Association. However, identification of SBIs is significantly impeded since discovery of them is an incidental finding: there are no International Classification of Diseases (ICD) codes for SBI, and it is generally not included in a patient's problem list or any structured EHR field. Concept extraction is therefore necessary for the identification of SBI-related findings from neuroimaging reports. We can infer a case of SBI based on acuity (chronic), location (left basal ganglia lacunar), and the number of infarcts (one) from the sentence "A chronic lacunar infarct is also noted".

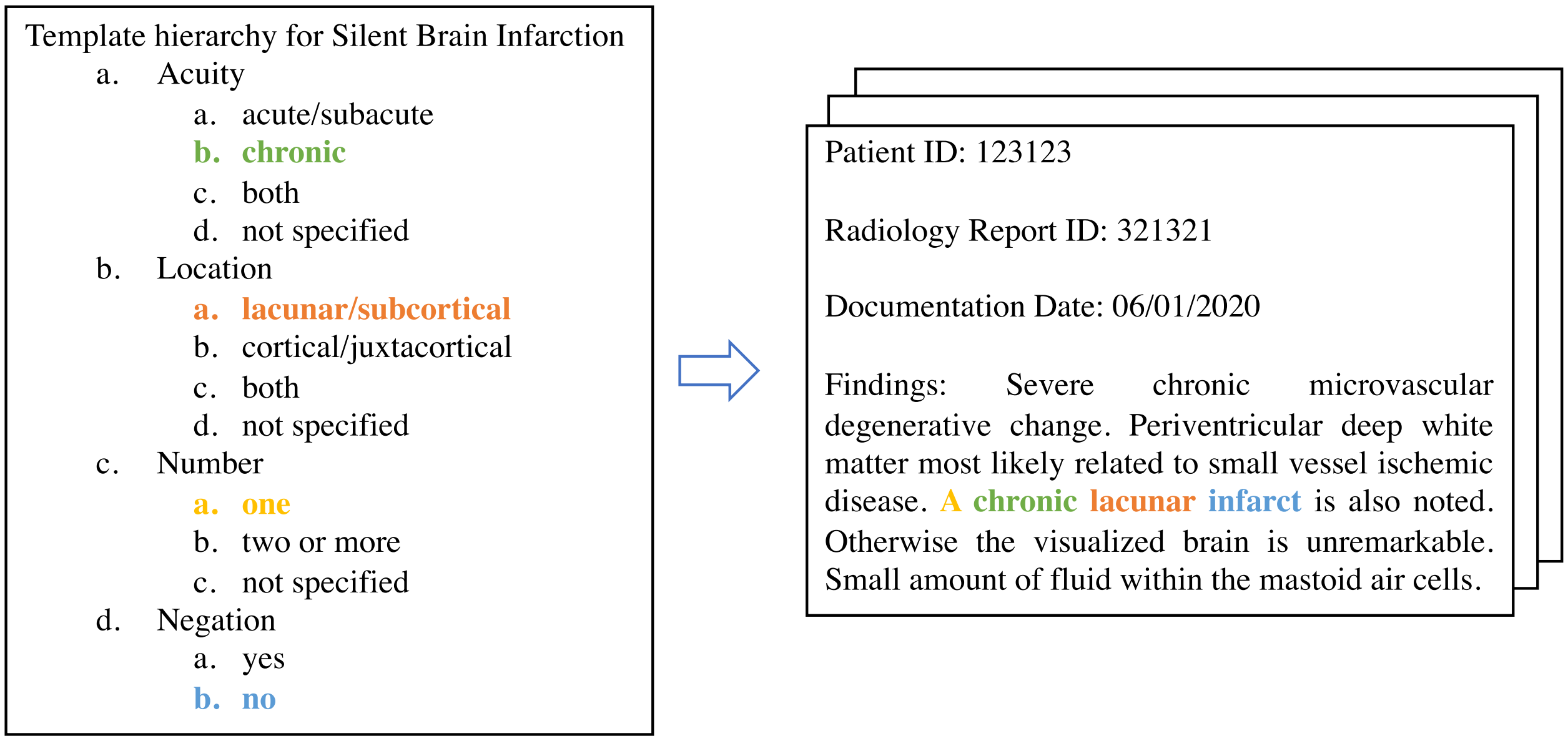

Figure 1. An example of using concept extraction for stroke research.

Methods for developing clinical concept extraction applications have been largely translated from the general NLP domain (4), and can typically be stratified into rule-based approaches and statistical approaches with four categories: rule-based, traditional machine learning (non-deep-learning variants), deep learning, or hybrid approaches. For example, an early attempt of clinical concept extraction, the Medical Language Processing project, was adapted from the Linguistic String Project aiming to extract symptoms, medications, and possible side effects from medical records (12, 13) leveraging a semantic lexicon and a large collection of rules. The rise of statistical NLP in the 1990s (14) and recent advances in deep learning technologies (15-17) have influenced the methods adopted for clinical concept extraction. Despite these advances, methods for clinical concept extraction are generally buried within the methods section of the literature due to the complexity and heterogeneity associated with EHR data and the diverse range of applications. No singular method has proven to be globally effective.

Here, we provide a review of the methodologies behind clinical concept extraction, cataloguing development processes, available methods and tools, and specific considerations when developing clinical concept extraction applications. This systematic review of the concept extraction task aims to address the following questions: 1) what are the keys steps involved in the development of a clinical concept extraction application; 2) what are the trends and associations of clinical concept extraction research over different approaches; and 3) what are the unresolved barriers, challenges and future directions.

## 2. Method

This review was conducted following a process compliant with the Preferred Reporting Items for Systematic Reviews and Meta-Analyses (PRISMA) guidelines (18). A literature search was conducted, retrieving EHR-based concept extraction articles that were written in English and published from January 2009 through June 2019. Literature databases surveyed included Ovid MEDLINE In-Process & Other Non-Indexed Citations, Ovid MEDLINE, Ovid EMBASE, Scopus, Web of Science, and the ACM Digital Library. The implementations of search patterns were consistent across the different databases. The search query was designed and implemented by an experienced librarian (LJP) as: ("clinic" or "clinical" or "electronic health record" or "electronic health records" or "electronic medical record" or "electronic medical records" or "electronic patient record" or "electronic patient records" or "EHR" or "EMR" or "EPR" or "ATR") AND ("information extraction" OR "concept extraction" OR "named entity extraction" OR "named entity recognition" OR "text mining" OR "natural language processing") AND (NOT "information retrieval"). A detailed description of the search strategies used is provided in the Appendix.

A total of 10,441 articles were retrieved from five libraries, of which 6,686 articles were found to be unique. The articles were then filtered manually based on the title, abstract, and method sections to keep articles with EHR-based clinical information extraction from English text. After this screening process, 928 articles were considered for subsequent categorization. We conducted additional manual review to keep articles with methodology description focusing on clinical concept extraction. The final inclusion criteria for the target papers are as follows 1) using concept extraction methods, 2) applied to EHR data in English, 3) providing a methodological contribution via: a) presenting novel methods for clinical concept extraction, including introducing a new model, data processing framework, NLP pipeline, etc., or b) applying existing methods to a new domain or task. Articles without full text or methology description were excluded. Following this screening process, 228 articles were selected and categorized based on the methods used. A comprehensive full-text review of all 228 studies was performed by the study team. During the full-text review, the following information were extracted manually: method type (e.g. rule based, machine learning), domain (e.g. disease study area), sub-domain, data type, performance (e.g. f1-score and AUC), and overall summary of methology. A flow chart of this article selection process is shown in Figure 2.

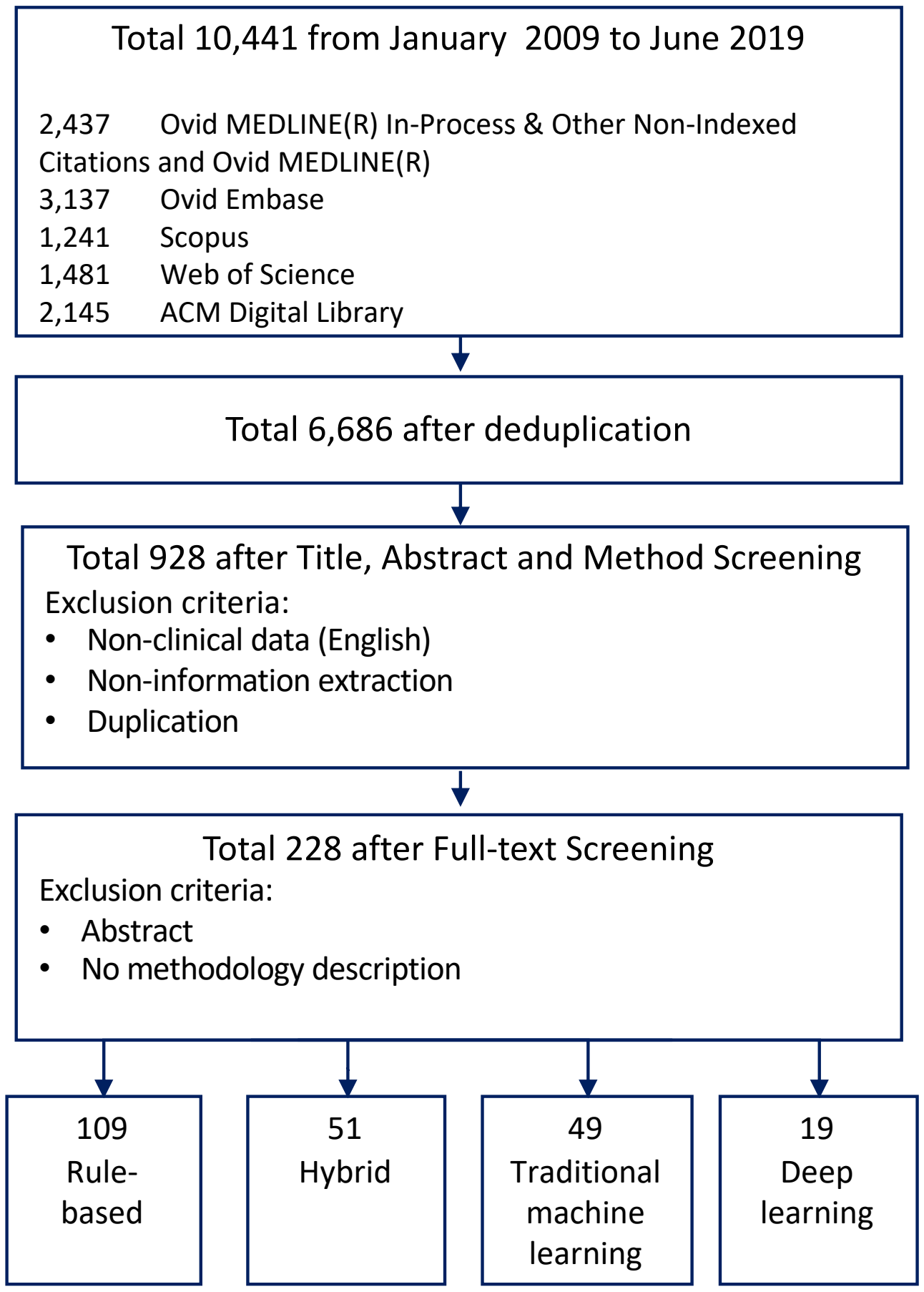

Figure 2. Overview of article selection process.

## 3. Results

Figure 3 shows the number of articles and associated methods available in our surveyed literature from January 2009 to June 2019. An upward trend in published clinical concept extraction research is observed, suggesting a strong need for harvesting information and knowledge from EHRs to support various clinical, operational, and research activities.

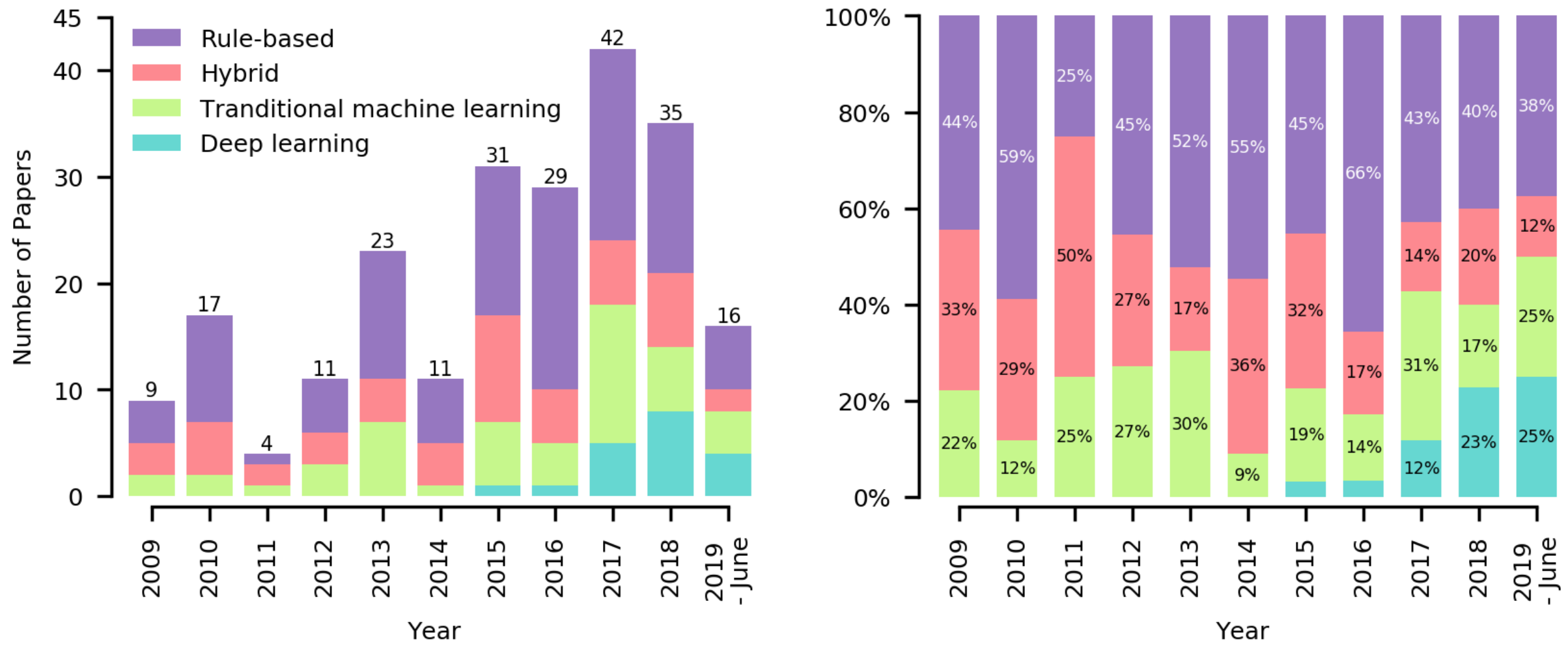

Figure 3. Trend view of clinical concept extraction research over different approaches.

A comprehensive mapping of each method to its associated application was illustrated in Figure 4 through a sunburst plot where the targets for concept extraction were categorized into four primary domains (i.e., diseases, drugs, clinical workflows, and social determinants of health) based on the focus of collected studies and clinical perspectives. The four domains were based on the focus of collected studies (frequency), clinical perspectives (importance) and the past clinical NLP shared tasks. We subsequently classified the four domains to more detailed sub-domains. The "Diseases" sub-domain was classified based on the ICD-9 classification system (19) given its stability and wide coverage in clinical practice. The sub-domain for clinical workflow optimization, drug and social behavior-related studies were determined by clinical perspectives and the uniqueness of the category.

Among the 228 surveyed articles, rule-based approaches were the most widely used approach for concept extraction (48%), followed by hybrid (22%), traditional machine learning (22%), and deep learning (8%). Despite the overall low utilization rate, adoption of deep learning techniques has increased drastically since 2017. We also observed that the three disease sub-domains with the highest coverage were (i) diseases of the circulatory system, neoplasms, and endocrine, (ii) nutritional and (iii) metabolic diseases, and immunity disorders. Six disease areas have the lowest coverage: (i) certain conditions originating in the perinatal period, (ii) complications of pregnancy, childbirth, and the puerperium, (iii) congenital anomalies, (iv) diseases of the skin and subcutaneous tissue, (v) diseases of the blood and blood-forming organs, and (vi) injury and poisoning. The popularity of the sub-domains was contributed to by many reasons. Particularly, we found shared-tasks can promote research activities in specific sub-domains. In addition, the disease sub-domains that cannot be addressed by disease classification codes alone were more likely to adopt concept extraction techniques. For example, incidental findings such as leukoaraiosis and asymptomatic lesions were reported to adopt concept extraction when ICD codes were not available (20, 21).

Under the clinical workflow domain, we observed that measurement, patient identification, and quality were broadly studied. The clinical note was the most utilized data type (78%), followed by radiology reports (13%) and pathology reports (3%). On the other hand, microbiology reports, claims, and operative reports were the least utilized data type.

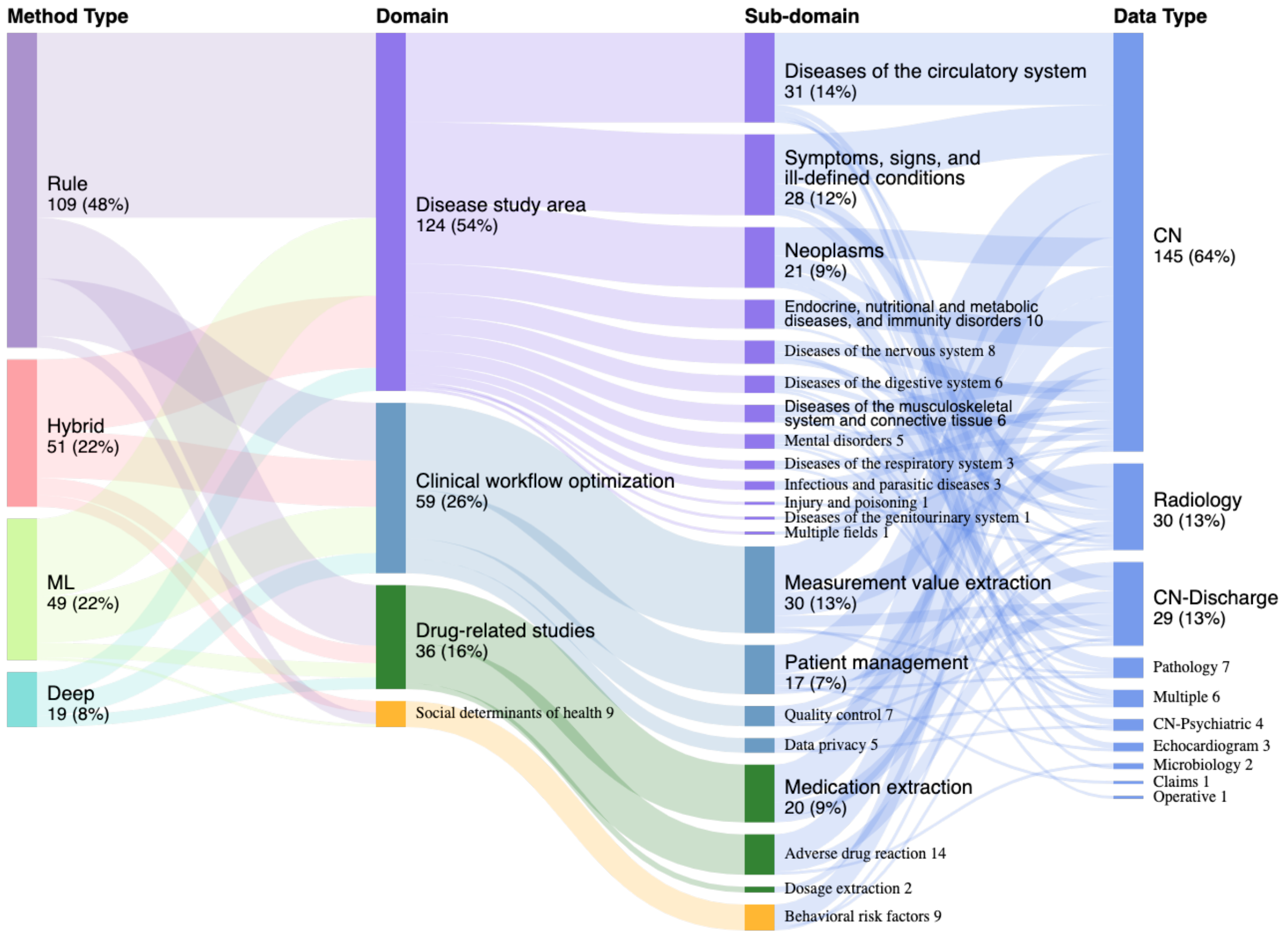

Figure 4. Overview of the method utilization.

Based on the performance of past shared tasks and evaluation on benchmarking datasets, we summarized the state of the art performance over each domain as of June 2019. The contextual pre-trained language models, such as Bidirectional Encoder Representations from Transformers (BERT), leveraged the bidirectional training of transformers, an attention mechanism that learns contextual relations between words, resulting in state-of-the-art results in six concept extraction tasks. Hybrid and traditional machine learning achieved the best performance on two tasks. In Table 1 we summarize a list of past shared tasks with the state-of-the-art methods.

Table 1. Benchmark of clinical concept extraction tasks

| **Domain** | **Task Description** | **F-measure** | **Method** | **Model** |
|---|---|---|---|---|
| Clinical workflow optimization | Automatic de-identification and identification of medical risk factors related to coronary artery disease in the narratives of longitudinal medical records of diabetic patients | 93.0 | Deep learning | BioBERT (22) |
| | i2b2 2006 1B de-identification: Automatic de-identification of personal health information | 94.6 | Deep learning | BioBERT (22) |
| Drug-related studies | i2b2 2010/VA : Medical problem extraction | 90.3 | Deep learning | BERT-large (23) |
| | i2b2 2009 medication challenge: Identification of medications, dosages, routes, frequencies, durations, and reasons | 85.7 | Hybrid | CRF, SVM, Context Engine(24) |
| Disease study area | ShARe/CLEFE 2013: Named entity recognition in clinical notes | 77.1 | Deep learning | BERT-base (P+M) (25) |
| | SemEval 2014 Task 7: Identification and normalization of diseases and disorders in clinical reports | 80.7 | Deep learning | BERT-large (23) |
| | SemEval 2014 Task 14: Named entity recognition and template slot filling for clinical texts | 81.7 | Deep learning | BERT-large (23) |
| Social determinants of health | I2B2 2006: Smoking identification challenge | 90.0 | Machine learning | SVM (26) |

The steps involved in the development of a clinical concept extraction application can be divided into three key components: 1) task formulation, 2) model development, and 3) experiment and evaluation (Figure 5). In the following subsections, we provide a detailed review of methods used for each of them. In section 3.1, we describe how to formulate a task in two different settings. Section 3.2 provides an overview of the approaches and summarizes the features for model development and the specific methods for concept extraction. Section 3.3 discusses methods for performing experiment and evaluation.

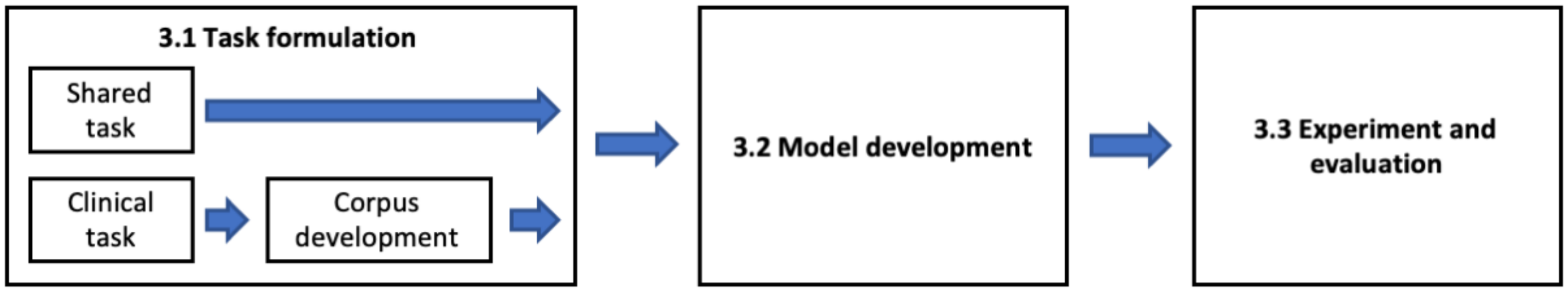

Figure 5. The development process of clinical concept extraction applications.

### 3.1 Task formulation

All studies were categorized into one of two experimental settings: shared-task and practice. The shared-task setting is defined as either participation in or usage of resources produced from a shared-task to conduct concept extraction-related research. The practice setting (e.g. General Internal Medicine, Orthopaedic Medicine) involves direct use of the EHR for concept extraction in real-world scenarios. Among a total of 228 articles, 63 (28%) were categorized as belonging to a shared task setting and 165 (72%) were categorized as belonging to a practice setting. The overall prevalence proportions for rule-based, hybrid, traditional machine learning, and deep learning were 35%, 32%, 18%, and 15% respectively in the shared task setting and 51%, 20%, 23%, and 6% respectively in the practice setting. A comparison between the two settings revealed a substantial difference in relative usage of statistical approaches and rule-based approaches: the practice setting had a much higher usage of rule-based approaches whereas deep learning and other statistical approaches had a higher prevalence of usage in the shared task setting.

Per the first author affiliations from 156 articles indexed in PubMed, 87 (56%) were affiliated with an academic institution, 50 (32%) with a medical center, 9 (6%) with an industry-based research institution, and 6 (4%) with a technology company. The remaining 3 authors (2%) were affiliated with a healthcare company, a medical library, and a pharmaceutical company.

**3.1.1 Shared-task setting** - Shared-tasks have successfully engaged NLP researchers in the advancement, adoption, and dissemination of novel NLP methods. Furthermore, because shared-task corpora are usually made accessible with well-defined evaluation mechanisms and public availability, they are usually regarded as standard benchmarks. Well-known tasks focusing on clinical concept extraction include the Informatics for Integrating Biology and the Bedside (i2b2) challenges (27-31), the Conference and Labs of the Evaluation Forum (CLEF) eHealth challenges (32, 33), the Semantic Evaluation (SemEval) challenges (32-34), BioCreative/OHNLP (35-38), and the National NLP Clinical Challenge (n2c2) (39).

The winning system in many past shared tasks typically used a hybrid approach, combining supervised traditional machine learning or deep learning for concept extraction with feature representation trained through unsupervised learning algorithms. The current state-of-the-art models use the long short-term memory (LSTM) (40) variant of bidirectional recurrent neural networks (BiRNN) with a subsequent conditional random field (CRF) decoding layer (15, 22, 23, 41). The convolution neural network (CNN)-based network, such as Gate-Relation Network (GRN), also performs well in tasks requiring long-term context information (42).

Recently, contextual pre-trained language models, such as Bidirectional Encoder Representations from Transformers (BERT), leverage bidirectional training of transformers, an attention mechanism that learns contextual relations between words, resulting in state-of-the-art results in the concept extraction task (15) (22) (23) (25).

**3.1.2 Practice setting** - Compared with the shared-task setting, the development of concept extraction applications in practice settings is more variant, costly, and time consuming. Tasks in practice settings can be much more specialized or ill-defined depending on the disease and use-case. Furthermore, there is an additional need for study teams to create the fully-annotated corpora themselves (whereas the corpora are provided in shared tasks), which can be expensive and time consuming. However, it is a necessary step to ensure rigorous evaluation of any developed technologies.

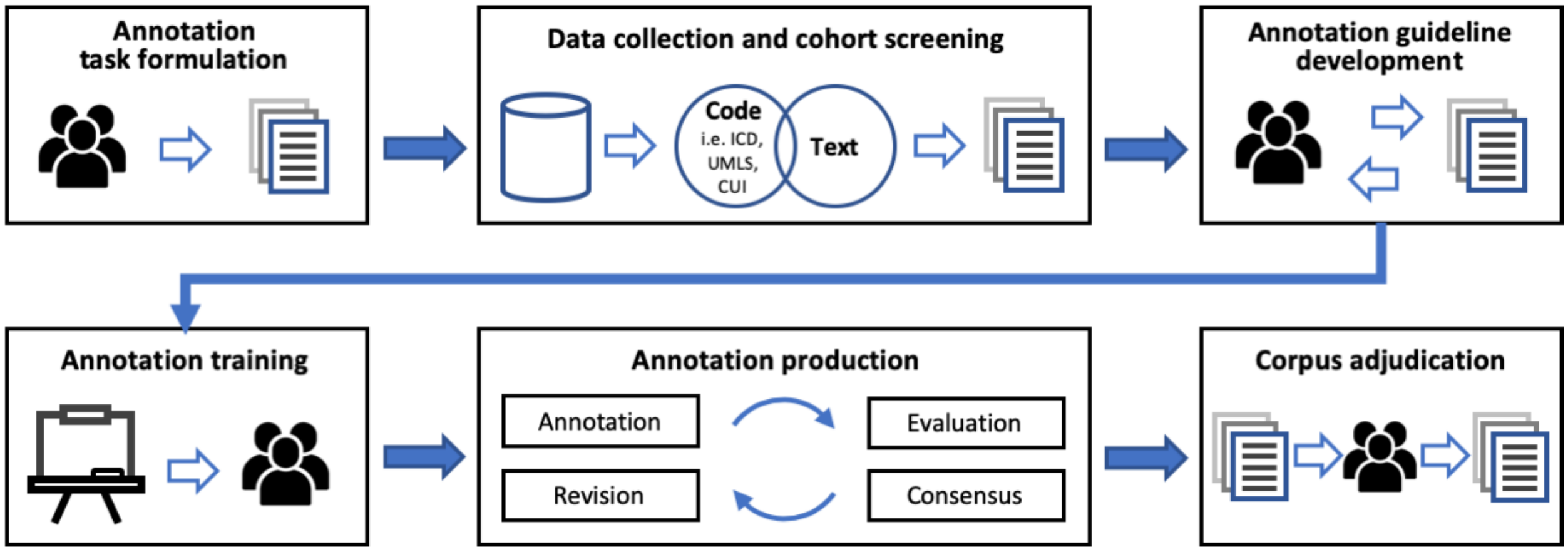

Figure 6. Data collection and corpus annotation process.

There are some variations in the process of creating gold standard corpora in practice settings due to the heterogeneity of data caused by variations in institutional processes and clinical workflows. Based on our review, we determined that the typical tasks involved in this process included annotation task formulation, data collection and cohort screening, annotation guideline development, annotation training, annotation production, and corpus adjudication (43-46) (Figure 6). Formulation of a task in a practice setting involved definition of points of interest (see Table 3 - Description), execution of a literature review, consultation with domain experts, and identification of study stakeholders such as abstractors and annotators with specialized knowledge. This step was then followed by definition and/or creation of a study population or cohort. For example, a concept extraction application of geriatric syndromes was developed based on the cohort of 18,341 people who were 1) 65 years or older, 2) received health insurance coverage between 2011 and 2013 and 3) were enrolled in a regional Medicare Advantage Health Maintenance Organization (47). Once the cohort definition was finalized, data was screened and retrieved. Studies have found the usefulness of leveraging open-source informatics technologies such as i2b2 or customized application programming interface and SQL queries for automatic screening and retrieval (48). Subsequent to dataset definition and creation, development of a detailed annotation guideline specifying the common conventions and standards was necessary, ensuring that definitions created are scientifically valid and robust. Notably, this step involves prototyping a baseline guideline, performing trial annotation runs, calculating inter-annotator agreement (IAA), and consensus building.

Determining the appropriate number of annotators and the size of corpus for annotation is critical and often challenged by the resources available. In general, the process requires at least two annotators with (clinical) domain expertise to independently perform the annotation (49-51). Having only one annotator in the study is not recommended since data validity and reliability cannot be measured and ensured (51). For multi-site studies, the process requires at least two annotators from both sides in order to help estimate inter-institutional and intra-institutional variation (52). Based on the analysis of 18 articles with exclusive discussion of gold standard development process, 4 (22%) studies used a single annotator, 10 (56%) studies reported of using two annotators and one adjudicator, 3 (17%) studies reported of using multiple annotator but did not specify the number, and 1 (6%) study with no mention of annotator. The median, minimal and maximal number of annotated clinical documents were 251, 100 and 8,321 respectively. Most studies choose 200 to 600 documents as the study data size. Among these, 30% to 60% were randomly sampled for double annotation and IAA assessment. Process iteration was used to save the annotation cost and increase effectiveness. For example, Mayer et al reported of having two annotators to perform the initial annotation on 15 documents for training and consensus development. During the second iteration, another 30 new documents were applied and IAA was calculated. The process was repeated until the IAA reached to 0.85 (53).

Proper training and education were utilized to reduce practice inconsistency and ensure a shared understanding of study design, objectives, concept definition, and informatics tools. The training materials generated included an initial annotation guideline walkthrough; demos and instructions on how to download, install, and use the annotation software; case studies; and practice annotations. Annotation production is typically organized into several iterations with a significant amount of overlap in the data annotated by each individual annotator to ensure the ability to determine IAA. Finally, in cases where annotators disagree, conflicts were adjudicated by subject matter experts. All the issues encountered during the gold standard creation process were documented. We encourage interested readers to read articles by Albright et al. (49) and South et al. (54) for more information on the standard annotation process for concept extraction.

### 3.2. Model Development

This section provides an overview of the specific approaches for model development and summarizes features used for concept extraction. Based on our review, there are five model development approaches which are summarized in Table 2 with relevant references.

Table 2. Comparison of model develoment approaches.

| **Processes** | **Examples** |
|---|---|
| A. Rule-based | (55-69) |
| 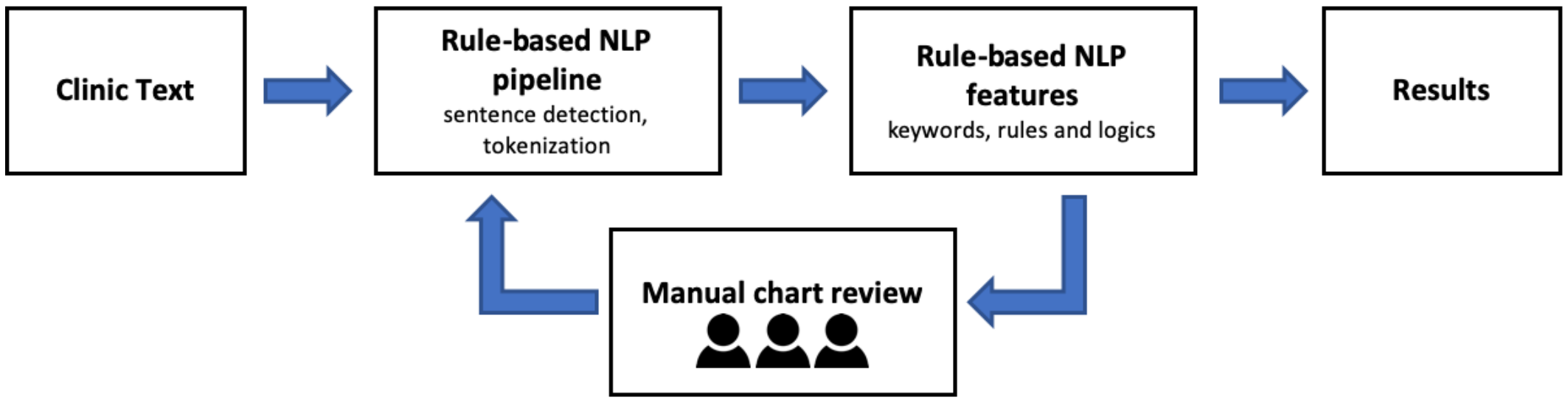  | |
| B. Traditional machine learning | (70-78) |
| 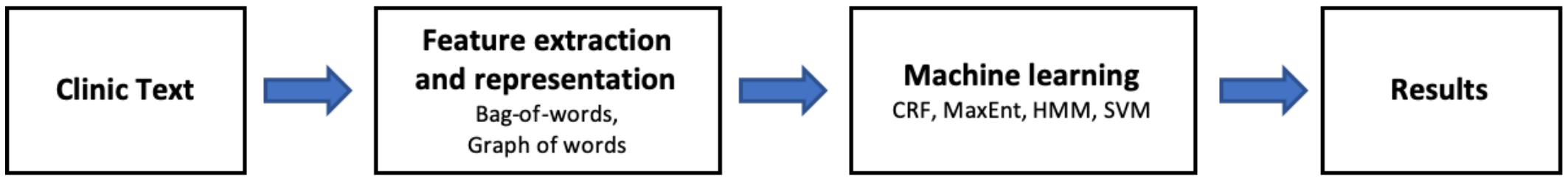  | |
| C. Deep learning | (79-85) |
| 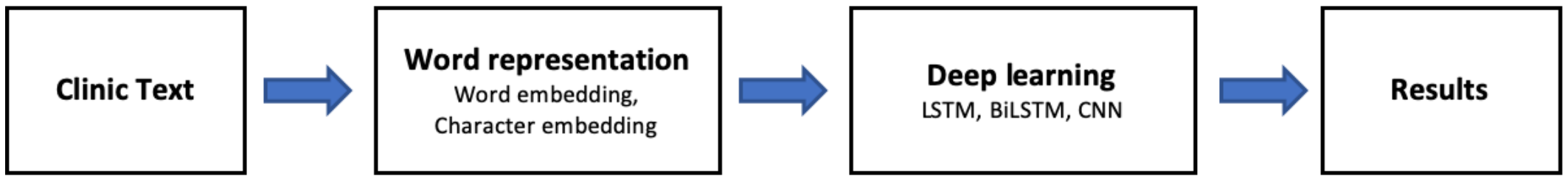  | |
| D. Terminal hybrid | (86-97) |
| 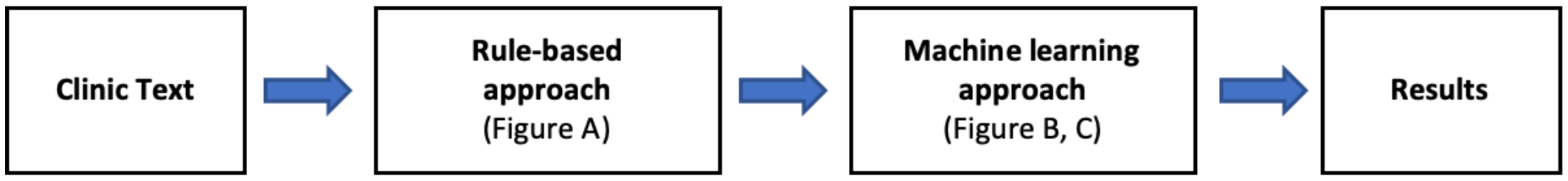  | |
| E. Supplemental hybrid | (98-105) |

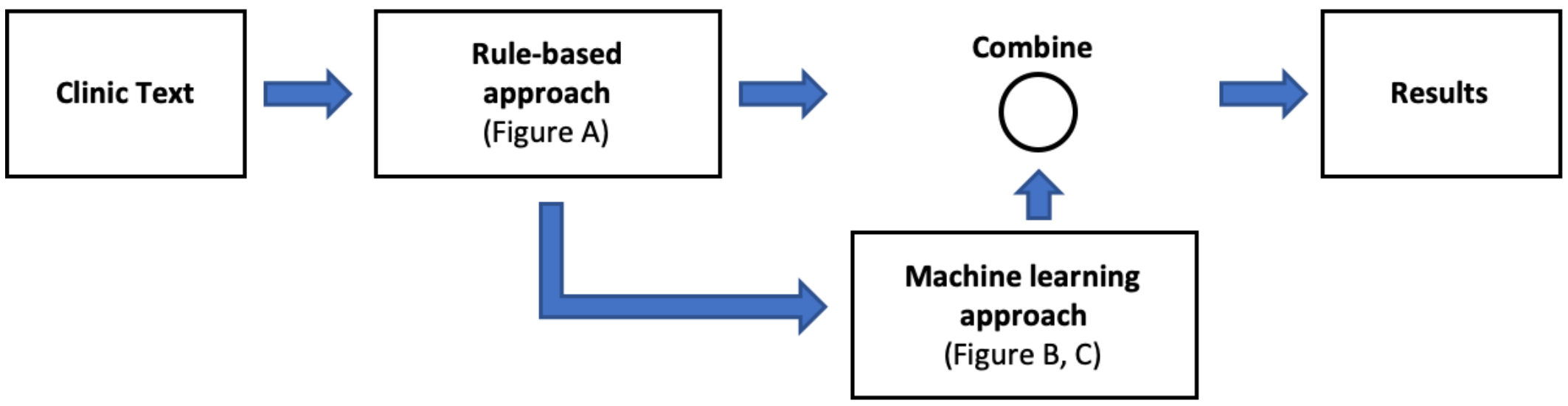

Designing computer systems to process text requires preparing the text in such a way that a computer can understand and perform operations on it. We broadly break these features down into categories of linguistic features, domain knowledge features, statistical features, and general document features, as summarized in Table 3. Features such as part of speech tagging, bag-of-words, and language models are used to give models more information on how to complete its task. Different approaches also tend to use different features. For example, the top three frequently used features for traditional machine learning and hybrid approaches are lexicon features (24%), syntactic features (20%), and ontologies (13%). The most intuitive text representation for the machine learning approach is a naïve one-hot encoding of the entire dictionary of words. Such a data representation presents several problems. First, there are over 180,000 words in the English dictionary, not including domain specific vocabulary (106). The vector space that the words are mapped to is both highly dimensional and extremely sparse, leading to inefficient and expensive computation. Second, such a representation assumes each "variable" is independent of the others, removing any semantic and synthetic features which may be beneficial for concept extraction tasks. For example, "cardiac" and "heart" would be represented by two independent variables, and therefore any concept extraction system could not easily learn their semantic similarities. Consequently, it is common to engineer data representations which encode semantic and syntactic similarities. Many rule-based systems are dependent on explicitly created features which tend to be more interpretable to humans. Alternatively, the latest techniques in deep learning create features which are often abstract (e.g. word embeddings or language models) and implicitly encode the semantic or synthetic features. Such features are very difficult to use by rule-based systems, but can greatly improve deep learning systems by reducing the dimensionality of the search space.

Table 1. An overview of features used for concept extraction.

| Feature Categories | Feature Types | Description | Features | Examples |
|---|---|---|---|---|
| Linguistic features | Lexicon-based feature | A dictionary or the vocabulary of a language | Bag-of-words (BoW); | "periprosthetic, joint, infection" (71, 77, 107) |
| | Orthographic features | A set of conventions for writing a language | Spelling; capitalization; hyphenation; punctuation; special characters | "igA"; "Pauci-immune vasculitis"; "BRCA1/2" (71, 108) |
| | Morphologic feature | Structure and formation of words | Prefix; suffix | "Omni" is the prefix of "Omnipaque" (91) |
| | Syntactic feature | Syntactic patterns presented in the text | Part-of-Speech, constituency parsing, dependency parsing, sequential representation (e.g. Inside–outside–beginning (IOB), B-beginning, I-intermediate, E-end, S-single word entity and O-outside (BIESO)) | "Communicating [verb] by [preposition] sinus tract [noun]" "Patient [O] has [O] an [O] acute [B] infarct [I] in [I] the [I] right [I] frontal [I] lobe [E]" (108, 109) |
| | Semantic feature | Semantic patterns presented in the text (whether a word is semantically related to the target words) | Synonym; hyponym/hypernym (etc.), frame semantics | "Loosing" is semantically related to "subsidence", "lucency", and "radiolucent lines" (68, 71) |
| Domain knowledge features | Conceptual feature | Semantic categories and relationships of words | Classifications and taxonomies, thesauri, ontologies: | ICD9/10, NCI Thesaurus, UMLS, UMLS Semantic Network, use case-specific code systems or controlled terminologies (71) |
| Statistical features | Graphic feature | Node and edge for each word in a document | Graph-of-Words (GOW) | Bi-directional representation: "grade" - "3", "3" - "grade" (110) |
| | Statistical corpus features | Features generated through basic statistical methods | Word length, TF-IDF, semantic similarity, distributional semantics, co-occurrence | "word1: parameter1, word2: parameter2" (70, 71) |
| | Vector-based representation of text | One-hot encoding<br>Word embedding<br>Sentence encoding<br>Paragraph encoding<br>Document encoding | Word2vec/doc2vec,<br>BERT,<br>one-hot character-level encoding. | Vector representations of text<br>(81, 82, 84) |
| General document features | Pattern and rule-based feature | A label for a note if certain rules satisfied | Logic (if–then) rules and expert systems | If "Metal" and "Polyethylene" then "Metal-on-Polyethylene bearing surfaces"<br>(58, 111) |
| | Contextual feature | Context information | Negated; status; hypothetical; experienced by someone other than the patient | "Patient [experiencer] does not [negation] have history of [status] infection"<br>(58, 59) |
| | Document structural feature | Structural and organizational patterns presented in the text and document | Section information; indentation; semi-structured information | "Family History [section]:<br>CVD<br>Diabetes<br>Hypertension"<br>(71, 112) |

Medical concept normalization is an essential process in order to take advantage of rich clinical information from large and complex free-form text. Medical concept normalization aims at mapping medical mentions to a corresponding medical concept in standardized ontologies such as Unified Medical Language System (UMLS) or controlled vocabularies such as SNOMED CT. The well-established clinical NLP tools such as MedLEE (113), MetaMap (114), cTAKES (115) and MedTagger (116) can achieve normalization using diverse approaches like a dictionary-lookup method and rule-based methods applying pattern recognition. In-depth studies explored specific medical concept normalization. For example, DNorm proposed a machine learning approach (pairwise learning-to-rank) to normalize mentions within a pre-defined lexicon (117). Hao et al. used heuristic rule-based approach to extract temporal expression and normalized them using the TIMEX standard (118). Lin et al. Proposed the strategy of contextual alias registry (CAR) and Chronological Order of Temporal Expressions (COTE) by focusing on domain-specific contextual alias dates and chronological order of clinical notes for temporal normalization (119). However, these mentioned-methods are not yet sufficient to autonomously deliver a comprehensive information of patient condition due to the complexity of clinical information and variations within medical domains (120). In sections 3.2.1. – 3.2.4., we are going to discuss the specific methods of applying feature extraction and normailization techniques for concept extraction.

**Rule-based approach**

Symbolic rule-based concept extraction methodologies use a comprehensive set of rules and keyword-based features to identify pre-defined patterns in text (58, 121). The rule-based approach has been adopted in many clinical applications due to their transparency and tractability, i.e. effectiveness of implementing domain specific knowledge. One particular advantage of using rule-based approaches is that the solution provides reliable results in a timely and low-cost manner, given the benefit of not needing to manually annotate a large amount of training examples (4). Based on specific tasks, the combination of rules and well-curated dictionaries can result in promising performance. Many tasks have used rule-based matching methods to varying levels of success (60, 122, 123). For example, in the 2014 i2b2/UTHealth de-identification challenge, the top four performing teams (including the winning team) used rule-based methodologies (124). In previous i2b2 challenges, the 2009 medication challenge reported 10 rule-based systems in the top 20 systems (30). In the i2b2/UTHealth Cardiac risk factors challenge, Cormack et al. demonstrated that a pattern matching system can achieve competitive performance with diverse lexical resources (61).

Ruleset development is an iterative process that requires manual effort to hand craft features based on clinical criteria, domain knowledge and/or expert opinions. The following steps were summarized based on the reported utilization and importance: (1) existing rule-based system adoption, (2) assessment of existing knowledge resources, and (3) development and refinement of features and logic rules.

First, the top performing rule-based applications often utilize existing NLP systems (frameworks) (56, 58, 60, 62, 125). There are many different NLP systems that have been developed at different institutions that are utilized to convert clinical narratives into structured data, including MedLEE (113), MetaMap (114), KnowledgeMap (126), cTAKES (115), HiTEX (127), and MedTagger (116). A comprehensive summary of NLP systems can be found in Wang et al (5). Second, adopting existing resources such as clinical criteria, guidelines, and clinical corpora can substantially reduce development efforts. The most common strategy is to leverage well curated clinical dictionaries and knowledge bases. The dictionary acts as a domain or task specific knowledge base, which can be easily modified, updated, and aggregated (128). Well-established medical terminologies and ontologies such as Unified Medical Language System (UMLS) Metathesaurus (129), Medical Subject Headings (MeSH) (130), and MEDLINE®, have been used as the basis for clinical information extraction tasks as they already contain well-defined concepts associated with multiple terms (131). This approach works best in tasks and situations where modifier detection and the recognition of complex dependencies in the document are not necessary (132, 133). Since the common features being exploited are morphologic, lexical and syntactic, dictionary-based methods are highly interpretable, adoptable, and customizable. In the 2014 i2b2/UTHealth challenge, Khalifa and Meystre leveraged UMLS dictionary lookup to identify cardiovascular risk factors and achieved an F-measure of 87.5% (60). Table 4 summarizes a list of dictionaries and knowledge bases that were used in our surveyed papers.

Table 2. An overview of the dictionaries and knowledge bases used for clinical concept extraction.

| Dictionary/Knowledge base | Description | Link | Examples |
| --- | --- | --- | --- |
| Unified Medical Language System (UMLS) | Biomedical thesaurus organized by concept and it links similar names for the same concept | https://www.nlm.nih.gov/research/umls/knowledge_sources/metathesaurus/ | (123, 134-137) |
| Wikipedia | PHI Category Description Source | https://en.wikipedia.org/wiki/ | (138) |
| MeSH (Medical Subject Headings) | MeSH is the controlled vocabulary based on PubMed articles indexing developed by NLM | https://www.ncbi.nlm.nih.gov/mesh | (122, 123, 139) |
| ICD-O-3 | International Classification of Diseases for Oncology | https://www.who.int/classifications/icd/adaptations/oncology/en/ | (140) |
| RadLex | Radiology lexicon | http://www.radlex.org/ | (123) |
| BodyParts3D | Anatomical concepts | https://dbarchive.biosciencedbc.jp/en/bodyparts3d/download.html | (123) |
| NCI Database | Cancer related information | https://cactus.nci.nih.gov/download/nci/ | (122) |
| Berman taxonomy | Tumor taxonomy | https://www.ncbi.nlm.nih.gov/pmc/articles/PMC535937/bin/1471-2407-4-88-S1.gz | (140) |
| CTCAE | Common Terminology Criteria for Adverse Events | https://ctep.cancer.gov/protocolDevelopment/electronic_applications/ctc.htm | (141) |

Third, in many situations, singularly relying on dictionaries cannot fully capture all the patterns necessary to completely capture a concept. Customized rules are created to address complex patterns. The creation of customized rules is an iterative process involving multiple subject matter experts. At each iteration, the rules are applied to a reference standard corpus, and its results are evaluated. Based on the evaluation performance, domain experts review false positive mentions with the given context to determine the reasons (e.g. family history, negated and hypothetical sentences). The false negative mentions are usually caused by missing keywords and negation errors, and can be addressed by refining existing rules or establishing new. This pattern was then repeated until it reached to a reasonable performance (e.g. maximal F-measure with balanced precision and recall). For example, Cormack et al. leveraged data-driven rule-based approach by starting with high recall rules and refining them to increase precision while maintaining recall with contextual patterns based on observations (61). Kelahan et al. extracted impression text and assigned positive and negative labels to sentences based on manual rules. The final label of the radiology report wass determined based on the frequency of sentence labels (64).

### 3.2.1. Traditional machine learning approach

Advances in methods have revitalized interest in statistical machine learning approaches for NLP, for which the non-deep-learning variants are typically referred to as “traditional” machine learning. Traditional machine learning is capable of learning patterns without explicit programming through learning the association of input data and labeled outputs (142-145). The learning function is inferred from the data, with the form of the function limited only by the assumptions made by the learning algorithm. Although feature engineering can be complex, the ability to process and learn from large document corpora greatly reduces the need to manually review documents and also has the possibility of developing more accurate models. However, in contrast to deep learning methods which learn from text in the sequential format in which it is stored, traditional machine learning approaches require more human intervention in the form of feature engineering (Table 3).

The process of developing traditional machine learning models can be summarized into the following steps: data pre-processing, feature extraction, modeling, optimization, and evaluation. Raw text is difficult for today’s computers to understand. In particular, non-deep learning methods were developed to learn from categorical or numerical data. Therefore, it is common to pre-process the text into a format that is readily computable. There are a wide variety of different pre-processing methods which have been proposed, however they are outside the scope of this manuscript. Standard pre-processing techniques include sentence segmentation that splits text into sentences, tokenization that divides a text or set of text into its individual words, stemming that reduces a word to its word stem, POS marking up a word in a text (corpus) as corresponding to a particular part of speech, and dependency parsing. The NLTK and the Stanford CoreNLP were the two most popular toolkits for performing data pre-processing (73, 76, 107, 146-148).

The majority of traditional machine learning approaches leverage a bag-of-words model for the word representation (70, 107, 109, 149-151). The bag-of-words model often tokenizes words into a sparse, high dimensional one-hot space. Although simple, this approach introduces sparsity, greatly increases the size of data, and also removes any sense of semantic similarity between words. Building on top of an existing bag of words feature, Yoon et al. proposed graph-of-words, a new text representation approach based on graph analytics which overcomes these limitations by modeling word co-occurrence (110). Chen et al. proposed a clustering method using Latent Dirichlet Allocation (LDA) to summarize sentences for feature representation (152). Another approach to text representation is to automatically learn abstract, low-dimensional representation of the words. Common approaches to this are continuous bag of words (CBOW) (79, 135) or Word2vec (81, 83-85). Recently, advanced embedding and language representations have further improved state-of-the-art clinical concept extraction. Word embeddings (153) capture latent syntactic and semantic similarities, but cannot incorporate context dependent semantics present at sentence or even more abstract levels. Peter et al. addressed this issue through training a neural language model which was able to capture the semantic roles of words in context. They found that the addition of the neural language model embeddings to word embeddings yielded state-of-the-art results for named-entity recognition and chunking (41). From a more abstract approach, Akbik et al. proposed character-level neural language modeling to capture not only the latent syntactic and semantic similarities between words, but also capture linguistic concepts such as sentences, subclauses, and sentiment (154). Many of these embeddings can be used in conjunction with others. For example, Liu et al. used both token-level and character-level word embeddings as the input layer (81). Choosing the appropriate embedding for the task can have large effects on the end model performance (81, 135).

The labeling for concept extraction is typically more complex compared to the standard classification or regression task. This is because entities in text are varying in length, location of text, and context. Commonly reported labeling for traditional machine learning include boundary detection-based classification and sequential labeling. Boundary detection aimed at detecting the boundaries of the target type of information. For example, the BIO tags use B for beginning, I for inside, and O for outside of a concept. Sequential labeling based extraction methods transforms each sentence into a sequence of tokens with a corresponding property or label. One particular advantage of sequential labeling is the consideration of the dependencies of the target information. Despite that, classification-based extraction is more commonly used than sequential labeling based extraction. From the review, 15 articles were found to utilize boundary detection-based classification approaches and 10 articles reported utilizing a sequential learning approach.

Frequently used traditional machine learning models for clinical concept extraction include conditional random fields (CRF) (155), the Support Vector Machine (SVM) (156), Structural Support Vector Machines (SSVMs) (157), Logistic Regression (LR) (158), the Bayesian model, and random forests (159). Among the aforementioned models, CRFs and the SVM are the two most popular models for clinical concept extraction (71). CRFs can be thought of as a generalization of LR for sequential data. SVMs use various kernels to transform data into a more easily discriminative hyperspace. Structural Support Vector Machines is an algorithm that combines the advantages of both CRFs and SVMs (71). When Tang et al. compared SSVMs and CRF using the data sets from 2010 i2b2 NLP challenge, the SSVMs achieved better performance than the CRFs using the same features (71). The summarized traditional machine learning approaches can be found in Table 5.

Table 3. Summary of traditional machine learning (non-deep learning) approaches.

| **Learning Task** | **Tag Examples** | **Word Representation** | **Model Examples** | **Examples** |
|---|---|---|---|---|
| Boundary detection-based classification | Word position tag (e.g. BIO, BIESO); Binary outcome tag (e.g. 0, 1) | Flat or naive representation; clustering-based word representation; distributional word representation | SVM; SSVM, Naïve Bayes; Decision Trees; AdaBoost; RandomForests; MIRA(160) | (55, 146, 161, 162) |
| Sequential learning | Word level tag (e.g. POS) | Sequential representation | CRF; Hidden Markov Model (HHM); Maximum Entropy | (73, 160, 163-165) |

| | |
|---|---|
| | Markov Models (MEMMs) |

### 3.2.2. Deep learning approach

Deep learning is a subfield of machine learning that focuses on automatic learning of features in multiple levels of abstract representations (17, 166, 167). The algorithms are largely focused around neural networks such as recurrent neural networks (RNN) (168-170), CNNs (42, 171-173) and transformers (174), although there are a few other niche approaches. Deep learning has led to revolutionary developments in many fields including computer vision (175, 176), robotics (177, 178), and NLP (17, 84, 179). In contrast to traditional machine learning paradigms, deep learning minimizes the need to engineer explicit data representations such as bag-of-words or n-grams.

Many of the deep learning applications in concept extraction have used either variants of RNNs or CNNs. CNNs rely on convolutional filters to capture spatial relationships in the inputs and pooling layers to minimize computational complexity. Although these have been found to be exceptionally useful for computer vision tasks, CNNs may have a difficult time capturing long distance relationships that are common in text (180). RNNs are neural networks which explicitly model connections along a sequence, making RNNs uniquely suited for tasks that require long-term dependency to be captured (181, 182). Conventional RNNs are, however, limited in modeling capability by the length of text (and therefore limited in the maximum distance between words) due to problems with vanishing gradients. Variants such as LSTM (40) and gated recurrent unit (GRU) (183) have been developed to address this issue by separating the propagation of the gradient and control of the propagation through “gates”. While shown to be effective, these only diminish the issue rather than completely solve it, being still limited to sequence lengths on the order of 10s-100s of words long (181). Furthermore, training these models is computationally intensive and difficult to parallelize as the weights need to be trained in series. Recently, the transformer architecture has been proposed to solve many of these problems. The transformer architecture circumvents the need to sequentially process text by processing the entire sequence at once through a set of matrix multiplications, allowing the network to memorize what element in the sequence is important (174). If the sequences are lengthy, the memory requirements for training are substantial. Breaking up the sequence into smaller pieces and adding subsequent layers is therefore needed to allow the model to accommodate long sequences of text without crippling memory constraints. Thus, transformers can effectively model relationships with long word distance and are much more computationally efficient compared to RNN variants. Models based on this architecture such as BERT (15) and GPT (184) have yielded significant improvements for state-of-the-art performance in many NLP tasks (25).

Meanwhile, using a deep learning model does not preclude using a traditional machine learning model. Although deep learning models are powerful feature extractors, other models may have specific attributes which suit particular needs of the problem. This is evident as many of the researchers combined conditional random field (CRF) with various deep neural networks (word embeddings input) to improve the performance on NER, such as Bi-LSTM-CRF (44%), Bi-LSTM-Attention-CRF (22%), and CNN-CRF (22%). This is to take advantage of their relative strengths: long distance modeling of RNNs and CRF's ability to jointly connect output tags. Choosing the correct algorithm that is appropriate for both the research setting and dataset is key to produce a successful model. Interested readers can refer to a recent review by Wu et al (17) for a more in-depth overview.

### 3.2.3. Hybrid approach

Hybrid approaches combine both rule- and machine learning-based approaches into one system, potentially offering the advantages of both and minimizing their respective weaknesses. There are two major hybrid approaches, as shown in Section 3.2 - Table 2. Depending on how traditional machine learning approaches were leveraged, we named these two architectures as either terminal hybrid approaches or supplemental hybrid approaches. In a terminal hybrid approach, rule-based systems are used for feature extraction, where the outputs became features used as input for the machine learning system, and the machine learning system is then a terminal step that selects optimal features. We found that the terminal hybrid approaches used in 31 out of 51 studies were in this category. For example, Wang and Akella (91) used NLP features, such as semantic, syntactic, and sequential features, as input to a supervised traditional machine learning model to extract disorder mentions from clinical notes. Other applications of hybrid systems include automatic de-identification of psychiatric notes (185, 186) and detection of clinical note sections (187). Section 3.2 - Table 3 lists the available NLP features used in the included studies.

In supplemental hybrid approaches, machine learning approaches are used to patch deficiencies in extraction of entities that have poor performance when extracted by purely rule-based approaches. In one study, such a supplemental hybrid system was incorporated with a user interface for interactive concept extraction (188). In another example, Meystre et al. (105) leveraged a traditional machine learning classifier to extract congestive heart failure medications as a supplement to the rule-based system that extracted mentions and values of left ventricular ejection fraction, amongst other concepts, for an assessment of treatment performance measures.

Using a hybrid approach may have the benefit of achieving high performance. Although traditional machine learning systems tend to do best when the task dataset has well-balanced outcomes, many concept extraction tasks' datasets are highly imbalanced, therefore making learning difficult. Farkas and Szarvas added specific "trigger words" as rules to improve their traditional machine learning de-identification system (189). Explicitly crafting the rules for these "trigger words" effectively created a "balanced" outcome, improving the algorithm's ability to correctly learn patterns. Yang and Garibaldi leveraged a dictionary-based method to supply a CRF model for medication concept recognition. The hybrid model was ranked fifth out of the 20 participating teams on the 2014 i2b2 challenge (101). In a study to automatically extract heart failure treatment performance metrics, the hybrid system outperformed both rule- and traditional machine learning-based approaches (190).

### 3.3. Experiment and Evaluation

Rigorously evaluating model performance is a crucial process for developing valid and reliable concept extraction applications. Evaluation is usually performed at one of several levels: a patient level, a document level, a concept level, or a defined episode level (with temporality). The specific level selected with which evaluation was performed was typically determined based on the specific task or application. For example, patient level detection may be sufficient if the task is to detect patients with a disease or a disease phenotype. On the other hand, identification of the time of presentation likely requires an evaluation with a finer level of granularity. Determination of an evaluation method, as with any concept extraction task, should be made in consultation with clinical subject matter experts and in the context of the application.

The evaluation can be performed by construction of a confusion matrix or a contingency table to derive error ratios. Commonly used ratios measure the number of true positives (the predicted label occurs in the gold-standard label set), false positives (the predicted label does not occur in the gold-standard label set), false negatives (the label occurs in the gold-standard set but was not a predicted label), and true negatives (the total number of occurrences that are not predicted to be a given label minus the false positives of that label). From these measures, the standardized evaluation metrics, including sensitivity or recall, specificity, precision, or positive predictive value (PPV), negative predictive value (NPV), and f-measure, can then be determined based on the error ratios. Because traditional machine learning or deep learning models also provide probabilities, performance can be evaluated using the area under the ROC curve (AUC) and the area under the precision-recall curve (PRAUC).

A majority of concept extraction methods are evaluated using the hold-out method, where the model is trained on training (and possible validation) sets and evaluated on the held-out test set. Cross validation (CV) can also be used to estimate the prediction error of a model by iteratively training part of the data and leaving the rest for testing. However, applying CV for error estimation on tuned models using CV may yield a biased result (191). The nested-cross-validation that uses two independent loops of CV for parameter tuning and error estimation can reduce the bias of the true error estimation (192).

For multiclass prediction or classification, micro-average and macro-average are two methods of weight prioritization. The micro-average calculates the score by aggregating the individual rates (e.g. true positive, false positive, and false negative). In the task of epilepsy phenotype extraction, Cui et al. reported the micro-average to reflect each individual class under an imbalanced category distribution (56). The macro-average calculates the metrics score (e.g. precision and recall) by each class first and then averages by the number of classes. In the evaluation of CEGS N-GRID 2016 shared task, macro-average was used to equally evaluate multiclass symptom severity (193).

In addition, mean squared error was also used for evaluating multiclass prediction or classification tasks. Another important aspect of evaluation, Cohen's kappa coefficient, is used to measure the inter- or intra-annotator agreement. Human annotators are imperfect, and therefore various problems may be more or less difficult to annotate. This can be due to fatigue, variation in interpretations, or annotator skill. This measure can be important to giving context on the performance of the model, as well as on the difficulty of the concept extraction task. An extensive summary of evaluation methods for clinical NLP can be found in Velupillai et al. (194).

## 4. Discussion

Given that methods for clinical concept extraction are generally buried within the methods section of the literature, here, we have provided a review of the methodologies behind clinical concept extraction applications based on a total of 228 articles. Our review aims to provide some guidance on method selection for clinical concept extraction tasks. However, the actual method adopted for a specific task can be impacted by five factors: data and resource availability, domain adaptation, model interpretability, system customizability, and practical implementation. In the following, we discuss each of them in detail.

**Data and resource availability** In clinical concept extraction tasks, the availability of data is a key consideration for determining the type of methods. For example, the amount of data used to train machine learning, specifically deep learning methods may impact the reliability and robustness of the model. Rule-based systems have less of a demand for training data and can be more appropriate when the data set is relatively small. Alternatively, when large amounts of labeled data are available and the task has a significant degree of ambiguity, machine learning approaches can be considered. From section 3.1, it is apparent that machine learning is used more than rule-based approaches in shared-task settings, possibly due in part to the readily available deidentified and annotated health-related data (e.g. MIMIC, i2b2 corpus). The availability of clinical and external subject matter expertise is another consideration. For the rule-based approaches, developing rules may require substantial manual effort such as iterative chart review and manual feature crafting. It may be challenging to involve clinicians to help with case validation and shared-decision making. However, large amounts of a priori knowledge about the domain and clinical problem (e.g. clear concept definition, well-documented abstraction protocol) and availability of clinical experts are favorable indicators for success of the rules-based approach.

**Domain adaptation** One way to leverage available resources is through transfer learning, an approach to reuse a model trained over a prior task as the pre-trained model for a new task, has been widely used in deep learning and NLP specifically (195, 196). A popular example of pre-trained models is BERT (15), which applies a transformer model (174) over huge narratives to generate a robust language representation model. Meanwhile, multi-concept learning plays an important role in learning tasks jointly and provides significant insights to accelerate domain adaptation. This learning strategy could also be referred as multitask learning, aiming to make use of features from multiple datasets or different data modalities to improve the generalization performance of models (197). In the biomedical domain specifically, some biomedical named entity recognition (BioNER) tasks were done by leveraging the multitask learning strategy. For example, based on 15 cross-domain biomedical datasets including the category of anatomy, chemical, disease, gene/protein, and species, Crichton et al. utilized a convolutional layer and a fully connected layer as a shared component and a task-specific component respectively to complete a BioNER task (198). In another study, building upon the embedding layers, Bi-LSTM layer and CRF layer, Wang et al. built a novel multitask model with the cross-sharing structure and ran it over heterogeneous gene, protein, and disease datasets to assist BioNER (199). Multimodal-based multitask learning is also a research direction to conduct prediction jointly incorporating different collected data modalities (e.g., claims data, free text notes, images, and genome sequences). For example, Weng et al. developed an innovative representation learning framework to co-learn features derived from patch-level pathological images, slide-level pathological images, pathology reports as well as case-level structured data (200). In addition, Nagpal developed different multitask models (i.e., Multitask Multilayer Perceptron, Recurrent Multitask Network, and Deep Multimodal Fusion Multitask Network) to jointly model different data modalities from EHRs, including ICD codes and unstructured clinical notes (201)

**Model interpretability** Considering that many models will eventually be implemented for clinical use, the evaluation of a successful model may not solely rely on performance, but also on the interpretability, which refers to the model's capability to explain why a certain decision is made (202). In a practice setting, these explanations may serve as important criteria for safety and bias evaluations, and are usually referred to as a key factor of "user trust" (203, 204). There are two categories of interpretability: intrinsic interpretability and post-hoc interpretability (205). Intrinsic interpretability emphasizes models with certain architectures that can be self-explained. Models with intrinsic interpretability include rule-based approaches, decision trees, linear models, and attention models (202). These models are widely used in the clinical domain compared to state-of-the-art deep learning approaches, due to their explanatory capabilities, the transparency of model components, and the interpretability of model parameters and hyperparameters (206). For example, Mowery et al. used regular expressions along with different semantic modifiers to conduct concept extraction of carotid stenosis findings. Based on the clinical definition of carotid disease, semantic patterns were organized into laterality (e.g. right, left), severity (e.g. critical or non-critical), and neurovascular anatomy (e.g. internal carotid artery) (112). The semantic pattern successfully captured important findings with high interpretability. With the advancement of deep learning models, there has been a surge of interest in interpreting black-box models through post-hoc interpretability. This can be achieved by creating an additional model to help independently interpret an existing model. Successful techniques such as a model-agnostic approach allow explanations to be generated without accessing the internal model. However, the trade-off between interpretability and model performance needs to be considered when deciding which method to use.

**System customizability** Customizability measures how easily each model can be adapted when a concept definition is changed or there is an update to clinical guidelines. The rule-based approaches allow the model to be modified and refined based on existing implementation. For example, Sohn et al. applied a refinement process when deploying an existing pattern matching algorithm to a different clinical site to achieve high performance (207). Davis et al. adopted four previously published algorithms for the identification of patients with multiple sclerosis using the rules combined with multiple features including ICD-9 codes, text keywords, and medication lists (125). The final updated algorithm was shared on PheKB, a publicly available phenotype knowledgebase for building and validating phenotype algorithms (208). Xu et al. leveraged existing algorithms from the eMERGE (electronic Medical Records and Genomics) network to identify patients with type 2 diabetes mellitus (209). Regarding traditional machine learning and deep learning models, it is difficult to make customizations explicitly due to their data-driven nature. In order to accommodate the models to specific clinical problems, incorporating knowledge-driven perspectives (e.g., sublanguage analysis and biomedical ontology) with the models are commonly adopted to customize the models implicitly. For example, Shen et al., combined surgical site infection features generated by sublanguage analysis with decision tree, random forest, and support vector machines to mine postsurgical complication patients automatically from unstructured clinical notes (210). Casteleiro et al., combined the Word2vec model with knowledge formalized in the cardiovascular disease ontology (CVDO) to provide a customized solution to extract more pertinent cardiovascular disease-related terms from biomedical literatures (211). The HPO2Vec+ framework provides a way to generate customized node embeddings for the Human Phenotype Ontology (HPO) based on different selections of knowledge repositories, in order to accelerate rare disease differential diagnosis by analyzing patient phenotypic characterization from clinical narratives (212).

**Practical implementation** When the models are carefully evaluated, they will be implemented for clinical and research use. The implementation process is highly dependent on institutional infrastructure, system requirements, data usage agreements, and research and practice objectives. A majority of the articles implemented the models into different standalone tools with the advantage of flexibility, low development cost, and low maintenance effort. For example, Fernandes et al. implemented a suicide ideation model by developing an additional platform to host the traditional machine learning component (213). Others have implemented the model into their institutional IT infrastructure, which includes an ETL process for document or HL7 message retrieval, a parser that does document pre-processing, an engine to host and run rule-based or traditional machine learning models, and a database to store extracted results (214).

**Open challenges** Beyond the aforementioned factors which impact the specific tasks, there are several open challenges which impact the overall field of clinical concept extraction. Like many AI tasks, reproducibility and system portability of proposed solutions are often in question. However, these questions are further challenged by the stringent privacy and security requirements posed by the regulations surrounding protected health information, thereby limiting ability to share data or produce fruitful collaborations (215). The lack of multi-institutional data is further exacerbated by the cost of annotating data, lack of detailed study protocols, and sometimes severe differences in EHR data between institutions (52). As the result, it has been shown that NLP algorithms developed in one institution for a study may not perform well when reused in the same institution or deployed to a different institution or for different studies. For example, Wagholikar et al. evaluated the performance of an NLP tagging system from two sites. The performance degrades when the tagger was ported to a different hospital (216). The differences in EHR systems, physician training, and data documentation standards are the source of significant clinical document variability and non-optimal performance of concept extraction systems.

Potential solutions such as federated learning have some advantages in dealing with data privacy issues (217). In a federated learning regime, several individual institutions (workers) collaborate to create a single model hosted at a centralized location (server) without the need to share data. Instead, the model is iteratively sent to the nodes to incrementally improve the results of the model. The trained models are then iteratively aggregated in the server. In this way, neither the server, nor other nodes have to see any data not residing at that specific node. However, this type of model development is limited to machine learning methods. Furthermore, the centralized model is typically developed using a centralized test set which as mentioned earlier, often does not approximate well to the data at individualized nodes.

**Limitations** There are some limitations in our study. First, this study may be biased and has the potential of missing relevant articles published due to the search strings and databases selected in this review. Secondly, we only included articles written in English with the focus on clinical information extraction. Articles written in other languages would also provide valuable information. Thirdly, our review did not include methods based on non-English EHRs. Finally, our study also suffers the inherent ambiguity associated with data element collection and normalization due to subjectivity introduced in the review process.

## Funding

This work was made possible by National Institute of Health (NIH) grant number 1U01TR002062-01 and R21AI142702.

## Competing Interests

The authors have no competing interests to declare.

## Contributors

SF performed the analysis and wrote the study. All authors participated in the study design, interpretation of the data, and contributed to manuscript editing and revisions. HH performed scientific visualization. HL conceptualized, designed, and edited the manuscript.

## Acknowledgements

We gratefully acknowledge Larry J. Prokop for implementing search strategies, and Katelyn Cordie and Luke Carlson for editorial support.